\documentclass[conference]{IEEEtran}
\usepackage{cite}
\usepackage{amsmath,amssymb,amsfonts}
\usepackage{algorithmic}
\usepackage{graphicx}
\usepackage{textcomp}
\usepackage{xcolor}
\usepackage{import}
\usepackage{siunitx}
\usepackage{mathtools}
\usepackage{subfigure}
\usepackage{booktabs}
\usepackage{multirow}
\usepackage{makecell}
\usepackage{tikz}
\usetikzlibrary{math}
\def\BibTeX{{\rm B\kern-.05em{\sc i\kern-.025em b}\kern-.08em
    T\kern-.1667em\lower.7ex\hbox{E}\kern-.125emX}}

\usepackage[absolute,overlay]{textpos}
\newcommand{\copyrightstatement}{
    \begin{textblock}{1.0}(0.096, 0.925)
         \noindent
         \footnotesize
         \copyright 2026 IEEE. Personal use of this material is permitted. Permission from IEEE must be obtained for all other uses, in any current or future media, including reprinting/republishing this material for advertising or promotional purposes, creating new collective works, for resale or redistribution to servers or lists, or reuse of any copyrighted component of this work in other works.
    \end{textblock}
}
\begin{document}
\copyrightstatement
\bstctlcite{IEEEexample:BSTcontrol}

\definecolor{mediumblue}{RGB}{0, 0, 205}
\definecolor{crimson}{RGB}{220, 20, 60}
\definecolor{salmon}{RGB}{250, 128, 114}

\title{Ring Mixing with Auxiliary Signal-to-Consistency-Error Ratio Loss for Unsupervised Denoising in Speech Separation}

\author{\IEEEauthorblockN{Matthew Maciejewski}
\IEEEauthorblockA{\textit{Human Language Technology Center of Excellence} \\
\textit{Johns Hopkins University}\\
Baltimore, USA \\
matt@mmaciejewski.com}
\and
\IEEEauthorblockN{Samuele Cornell}
\IEEEauthorblockA{\textit{Language Technologies Institute} \\
\textit{Carnegie Mellon University}\\
Pittsburgh, USA \\
samuele.cornell@ieee.org}
}

\maketitle

\begin{abstract}
Noisy speech separation systems are typically trained on fully-synthetic mixtures, limiting generalization to real-world scenarios. Though training on mixtures of in-domain (thus often noisy) speech is possible, we show that this leads to undesirable optima where mixture noise is retained in the estimates, due to the inseparability of the background noises and the loss function's symmetry. To address this, we propose ring mixing, a batch strategy of using each source in two mixtures, alongside a new Signal-to-Consistency-Error Ratio~(SCER) auxiliary loss penalizing inconsistent estimates of the same source from different mixtures, breaking symmetry and incentivizing denoising. On a WHAM!-based benchmark, our method can reduce residual noise by upwards of half, effectively learning to denoise from only noisy recordings. This opens the door to training more generalizable systems using in-the-wild data, which we demonstrate via systems trained using naturally-noisy speech from VoxCeleb.
\end{abstract}


\section{Introduction}
\label{sec:introduction}
Speech separation is the task of producing individual waveforms for each talker in a recording where multiple people have spoken at the same time.
With the advent of deep learning, the performance of speech separation systems has improved drastically, with many systems achieving Scale-Invariant Signal-to-Distortion Ratio~(SI-SDR)~\cite{sisdr} improvements of over \SI{20}{dB}~\cite{tfgridnet2} on the \SI{8}{kHz} wsj0-2mix~\cite{wsj2mix} dataset.
With errors now largely imperceptible to human listeners, the field has shifted toward more challenging conditions: moving beyond the narrow-band studio-quality read speech of wsj0-2mix to address the noisy, overlapping speech typical of real-world conversations~\cite{conversation_overlap}.

To broaden the conditions available to the research community, a number of additional speech separation datasets have been released, such as WHAM!~\cite{wham}, WHAMR!~\cite{whamr}, SMS-WSJ~\cite{sms_wsj}, and LibriMix~\cite{librimix}, which introduce noise and reverberation.
However, one notable property of these datasets is that the source speech is always single-talker studio-quality read speech, digitally summed to create speech mixtures, with the interfering noise and reverberation added artificially.
This is a potentially significant downside, as deep learning models are known to suffer from generalization issues that are best addressed by training on data from the same domain as the intended deployment conditions~\cite{recht2019imagenet, noisy_oracle, mixit, cornell2025recent}.
But, this reliance on synthetic data is not without reason: supervised neural network-based separation systems require paired training data with clean speech targets~\cite{noisy_oracle}, and obtaining such targets from real recordings would require solving the task the system is meant to learn.

A direction explored in a number of works is to mitigate this constraint through weakly or self-supervised approaches that would allow training on natural speech recordings.
For example, various approaches to unsupervised denoising have been explored~\cite{fujimura2021noisy,tzinis2022remixit,unse}, but these cannot be trivially extended to multi-talker separation.
A number of works have looked to spatial cues to perform unsupervised or semi-supervised separation~\cite{aralikatti2023reverberation,wang2023unssor,saijo2024enhanced}, but they generally do not address denoising and require multi-microphone recordings.
Among these multi-channel approaches only Neural FCASA~\cite{bando2023neural} \textit{does} perform joint denoising and separation but requires speaker activity labels as additional supervision.

Regarding single-channel approaches, the closest work to ours is Maciejewski et al.~\cite{noisy_oracle}, who study the effects of training with artificial mixtures of naturally-noisy speech, demonstrating a large performance gap compared to clean-target training.
However, their method does not explicitly attempt denoising and shows limited and inconsistent performance gains.
MixIT~\cite{mixit} and similar separation works~\cite{zhang2021teacher, karamatli2022mixcycle, saijo2023remixing} are motivated by the absence of isolated source recordings, training on synthetic mixtures of real mixtures to recover all constituent sources, and seem to naturally extend to this formulation: mixtures of noisy speech could be considered mixtures of mixtures of speech and noise signals that are to be independently estimated.
However, this extension of MixIT is also not trivial, and would still be subject to the issues of noise inseparability we demonstrate in Section~\ref{sec:proposed_method}.

To our knowledge, no single-channel speech separation system trained on noisy speech learns to reliably denoise without clean-speech supervision or, in other words, can perform unsupervised denoising for ``in-the-wild'' speech separation.

In this work, we specifically address this problem and analyze the effects of noisy-target training on the typical SI-SDR loss, showing that it results in an undesirable optimization minimum that includes noise in the output.
Then, we propose ring mixing, a batch construction strategy, alongside a new Signal-to-Consistency-Error Ratio~(SCER) auxiliary loss to compensate for this shortcoming.
Our method significantly reduces output noise with minimal impact on separation of speech, boosts generalization, and is capable of achieving noise removal on par with fully-supervised systems.

\section{Proposed Method}
\label{sec:proposed_method}
\subsection{Problem Formulation}
\label{ssec:problem_formulation}

In the most basic of conventional supervised speech separation systems, the problem is formulated as estimating two speech signals from their mixture.
Audio recordings generally hold the superposition principle, meaning the natural mixture of two audio signals can be approximated by simply adding the two waveforms together.
The system is then trained by summing single-talker recordings, inputting their mixture to the network, and training it to output the original recordings.

When moving to the noisy speech separation task, the formulation becomes more complicated.
The typical approach is to simulate a noisy mixture waveform $x \in \mathbb{R}^T$ by summing clean speech waveforms $s^\text{clean}_k \in \mathbb{R}^T, k\in\{1, 2\}$ and an additional noise waveform $n \in \mathbb{R}^T$:
\begin{align}
x = s^\text{clean}_1 + s^\text{clean}_2 + n \text{,}\label{eq:enh_sep}
\end{align}
where the network is trained in a supervised manner to output estimates $\hat{s}_k$ of $s^\text{clean}_k$, i.e. to both separate and denoise jointly.
A downside of this approach is that it requires the single-speaker speech recordings to be noiseless.
This greatly restricts the available pool of data to studio-quality speech and also eliminates the potential for in-domain training for realistic deployment scenarios.

An alternative approach, then, that allows the use of naturally-noisy (and potentially in-domain) speech, is to simply mix real noisy speech signals $s^\text{noisy}_k \in \mathbb{R}^T$ without artificially adding noise:
\begin{align}
x = s^\text{noisy}_1 + s^\text{noisy}_2 \text{,}\label{eq:noisy_sep}
\end{align}
where $s^\text{noisy}_k$ can be considered a sum of two unavailable underlying speech and noise signals: $s_k + n_k$.
Accordingly, the network is trained to produce estimates $\hat{s}_k$ of the noisy speech $s^\text{noisy}_k$, as the underlying clean speech signal $s_k$ is unavailable for supervision.
As such, it is \textit{not} being explicitly trained to denoise, and it is nontrivial to know what noise may or may not be included in estimates produced by a well-trained model.

Even setting this aside, however, this approach also has its own flaws.
One is that the amount of noise in the input mixture $x$ will be twice as much ($n_1+n_2$) as the ground truth recordings $s^\text{noisy}_k$.
A larger issue is that the task itself is somewhat malformed:
When a network is trained to separate $s_1^\text{noisy}$ from $s_2^\text{noisy}$, it is not only being trained to separate speech (discriminate $s_1$ from $s_2$) and denoise ($s_1$ from $n_2$), but also separate noise from noise ($n_1$ from $n_2$).
While the first two are generally possible due to the structure of speech, disentangling noise mixtures may be difficult or impossible depending on the type of noise.
Additionally, there is a largely unsolvable permutation problem, where even if the network \textit{can} discriminate all four signals, it must also correctly identify which noise signals pair with which speech signals. (In other words, outputting $s_1+n_2$ and $s_2+n_1$ is incorrect, even though there may be no information in the signals to indicate that $n_1$---not $n_2$---is the correct noise to pair with $s_1$.)

The goal of this work is to explore the noisy-target data formulation posed in equation~\eqref{eq:noisy_sep}.
We aim to analyze and demonstrate the consequences of using the typical supervised learning approaches with this data formulation, compared to the clean-target formulation of equation~\eqref{eq:enh_sep}.
Additionally, we aim to develop a new training scheme where the estimates $\hat{s}_k$ produced by a model using the noisy-target formulation are incentivized to estimate the underlying clean speech signal $s_k$ contained within $s_k^\text{noisy}$ even without direct supervision---that is, to learn joint separation \textit{and} denoising using only synthetic mixtures of noisy speech, with no clean speech supervision.

For clarity of notation, we will explicitly use $s_k^\text{clean}$ and $s_k^\text{noisy}$ in cases where we are referring to a specific type of ground truth supervision, and omit the superscript when variables are used in a more general sense.

\subsection{Issue with Noisy-Target Conventional Supervised Training}
\label{ssec:issue_with_conventional_supervision}

The typical loss function used to train networks is the negative of the standard Scale-Invariant Signal-to-Distortion Ratio~(SI-SDR)~\cite{sisdr} evaluation metric.
Setting aside the scaling issues for now, the per-source optimization objective for an estimate $\hat{s}_k$ of a ground truth signal $s_k$ is equivalent to:
\begin{equation}
\ell_\text{SDR}(\hat{s}_k; s_k) = -10\log\frac{||s_k||^2}{|| s_k - \hat{s}_k ||^2}. \label{eq:sdr_loss}
\end{equation}
Although this is formulated as an energy ratio in decibels in the style of a traditional Signal-to-Noise Ratio~(SNR), from an optimization perspective this is simply the logarithm of the mean squared error between $\hat{s}_k$ and $s_k$.

We would like to evaluate the behavior of $\hat{s}_k$ when $\ell_\text{SDR}$ is used under noisy-target supervision $s_k^\text{noisy}$.
Naively, the obvious solution would be that $\hat{s}_k \approx s_k+n_k$, i.e. would simply become the supervision target $s^\text{noisy}_k$.
However, there is evidence that networks cannot separate mixtures of background noise~\cite{noisy_oracle}, and so for example $\hat{s}_1$ could not be $s_1+n_1$, as the network would have had to separate $n_1$ from $n_2$.

To account for this, we develop a set $\mathcal{S}$ of potential estimates $\tilde{s}_k$ that could be produced by an optimally-performing network that is capable of perfectly estimating all signals except $n_1$ and $n_2$, which are assumed inseparable, and can only estimate their mixture $n_1+n_2$:
\begin{equation}
\tilde{s}_k \in \mathcal{S} = \{s_k + \lambda (n_1+n_2)\ |\ \lambda \in [0, 1]\} \text{.}\label{eq:set}
\end{equation}
This set represents the class of solutions where the speech signal $s_k$ has been estimated perfectly, the interfering speech signal has been entirely removed, and the network can freely adjust how much of the noise mixture $n_1+n_2$ is included in the estimate, ranging from entirely removed ($\lambda=0$) to entirely kept ($\lambda=1$) and everything in between.
One could consider separate $\lambda$ values $\lambda_1$ and $\lambda_2$ for $\hat{s}_1$ and $\hat{s}_2$, but given that there is no inherent distinction between source 1 and 2 (necessitating the use of Permutation Invariant Training~(PIT)~\cite{pit}), we claim that networks will treat each source the same and would accordingly have identical behavior for both sources.
We also note that the underlying assumptions behind $\mathcal{S}$ are supported experimentally in Section~\ref{sec:results_and_discussion}.

We then consider which value of $\lambda$ minimizes the $\ell_\text{SDR}$ loss.
This identifies the amount of residual noise incentivized by the loss under the noisy-target formulation and the relevant assumptions:
\begin{align}
\lambda^* & = \arg\min_\lambda \left[\ell_\text{SDR}(\tilde{s}_1, s_1^\text{noisy})+\ell_\text{SDR}(\tilde{s}_2, s_2^\text{noisy})\right] \\
& = \arg\min_\lambda \left[ \log||s_1^\text{noisy}-\tilde{s}_1||^2 + \log||s_2^\text{noisy}-\tilde{s}_2||^2\right] \\
\begin{split}
& = \arg\min_\lambda \hspace{1.65pt} \Bigl[ \log|| (1-\lambda)n_1 - \lambda n_2 ||^2 \\
& \hspace{5.5em} + \log|| (1-\lambda)n_2 - \lambda n_1 ||^2 \Bigr] \text{.}
\end{split}
\end{align}
We next expand the vector norms, relying on the common assumption~\cite{sisdr,bss_eval} that $\langle n_1, n_2 \rangle \approx 0$, which follows from audio being zero-mean and the recordings being independent:
\begin{align}
\begin{split}
\lambda^* & = \arg\min_\lambda \Bigl[ \log \left[ (1-\lambda)^2 ||n_1||^2 + \lambda^2 ||n_2||^2 \right] \\
& \hspace{5.5em} + \log \left[ \lambda^2 ||n_1||^2 + (1-\lambda)^2 ||n_2||^2 \right] \Bigr] \text{.}
\end{split}
\end{align}

While a closed-form solution for $\lambda^*$ is difficult to derive due to the summation within the logarithm, we instead highlight a few aspects of the function evaluated by the $\arg\min$:
\begin{itemize}
\item It is symmetric about $\lambda=0.5$.
\item If $||n_1||^2 = ||n_2||^2$, the minimum is at $ \lambda^* = 0.5$.
\item If $||n_1||^2$ or $||n_2||^2$ is $0$, the dual minima are $ \lambda^* \in \{0, 1\}$.
\end{itemize}
This is enough to support a useful characterization of the function:
If the amount of noise in each recording is roughly the same, the optimal value $\lambda^*$ is $0.5$.
As the total noise starts to become dominated by one recording, the optimum splits and drifts away from $0.5$ to dual optima $\epsilon$ and $1-\epsilon$ for $\epsilon \in (0, 0.5)$.

The intuition for this behavior should become more clear when one remembers that the magnitude of a vector is invariant to sign changes.
If $n_1$ is omitted from an estimate of $s_1^\text{noisy}$, the error term $s_1^\text{noisy}-s_1 = n_1$ has the same magnitude as if an estimate of $s_2^\text{noisy}$ includes $n_1$ erroneously: $s_2^\text{noisy}-(s_2+n_1+n_2) = -n_1$.
Since both sources are considered in the loss, there is an inherent symmetry---estimates with ``too much'' noise produce identical loss as estimates with ``too little'' noise, with the optimal value generally being to include half of the total noise in each estimate.

We emphasize that in all cases, networks are \textit{actively encouraged} to retain noise in the estimates, specifically at half amplitude ($\lambda^* = 0.5$) when the noise levels are balanced (very likely in cases where the data is all drawn from a single target domain).
Using unbalanced noise for a $\lambda = \epsilon$ minimum near $0$ might seem promising, but even ignoring potential challenges in data construction, we have generally observed that during training, networks first reconstruct the mixture, then suppress interferences, and are thus likely to get ``stuck'' in the $\lambda=1-\epsilon$ minimum instead, i.e. including the full noise in each estimate.

\subsection{Proposed Solution}
\label{ssec:proposed_solution}

In short, our solution is to break the symmetry where overestimating the noise in $s^\text{noisy}_1$ with the inclusion of $n_2$ results in the same error as underestimating by omitting $n_1$.
We do this by using the same single-talker recordings in multiple mixtures and enforcing consistency between their estimates.
In this case, underestimating by $n_1$ will \textit{not} result in a consistency penalty, but overestimating by $n_2$ \textit{will} result in a consistency penalty, as the two $n_2$ signals from different mixtures will not be the same.

\begin{figure}[t]
\centering
\subfigure[Normal batch]{
\begin{tikzpicture}
\tikzmath{
\colwidth = 20;
\rowwidth = 25;
\seglen = 30;
}
% mixtures
\foreach \i in {0,...,5} {
  \pgfmathtruncatemacro{\y}{floor(\i/2) * -\rowwidth};
  \pgfmathtruncatemacro{\xl}{Mod(\i, 2)*(\colwidth+\seglen)};
  \pgfmathtruncatemacro{\xr}{\xl+\seglen};
  \draw[crimson, line width=2pt](\xl pt, \y pt) -- (\xr pt, \y pt);
  \pgfmathtruncatemacro{\xm}{\xl+(\seglen/2)};
  \pgfmathtruncatemacro{\y}{\y+6};
  \pgfmathtruncatemacro{\idx}{\i+1}
  \node at (\xm pt, \y pt) {$x_{\idx}$};
}
% sources
\foreach \i in {0,...,11} {
  \pgfmathtruncatemacro{\y}{floor(\i/ 4) * -\rowwidth};
  \pgfmathtruncatemacro{\x}{Mod(\i, 4)*\seglen - floor(Mod(\i, 4)/2)*(\seglen-\colwidth)};
  \filldraw[mediumblue](\x pt, \y pt) circle (2.5pt);
  \pgfmathtruncatemacro{\y}{\y+8};
  \pgfmathtruncatemacro{\idx}{\i+1}
  \node at (\x pt, \y pt) {$s_{\idx}$};
}
\path(current bounding box.south) ++(0,-6pt);
\end{tikzpicture}
\label{sfig:normal_batch}
}
\subfigure[Ring mixing]{
\begin{tikzpicture}
\tikzmath{
\radius = 25;
}
\draw[color=crimson, line width=2pt](0, 0) circle (\radius pt);
\foreach \i in {0,...,5}
{
  % dots
  \pgfmathtruncatemacro{\x}{\radius * cos(\i * 60)}
  \pgfmathtruncatemacro{\y}{(\radius+1) * sin(\i * 60)}
  \filldraw[mediumblue](\x pt, \y pt) circle (2.5pt);
  % source labels
  \pgfmathtruncatemacro{\idx}{\i+1}
  \pgfmathtruncatemacro{\x}{(\radius+8) * cos(\i * 60)}
  \pgfmathtruncatemacro{\y}{(\radius+8) * sin(\i * 60)}
  \node at (\x pt, \y pt) {$s_{\idx}$};
  % mix labels
  \pgfmathtruncatemacro{\idx}{Mod(\i+1, 6)+1}
  \pgfmathtruncatemacro{\x}{(\radius+8) * -sin(\i * 60)}
  \pgfmathtruncatemacro{\y}{(\radius+6) * cos(\i * 60)}
  \node at (\x pt, \y pt) {$x_{\idx}$};
}
\end{tikzpicture}
\label{sfig:ring_mixing}
}
\caption{A 6-mixture batch with normal and ring mixing.}
\vspace{-8pt}
\label{fig:ring_diagram}
\end{figure}

More precisely, when generating data, we consider the batch in aggregate.
As the computation graph is no longer independent across training samples, we slightly shift our notation to accommodate all mixtures and sources:
Rather than each mixture $x$ being a summation of two sources $s_1$ and $s_2$, we will index both the mixtures and sources in the overall batch.
For example, in typical batch of $K$ mixtures, the first mixture $x_1$ would be the summation of the first two sources $s_1$ and $s_2$, the second mixture $x_2$ would be the summation of the next two sources $s_3$ and $s_4$, so on and so forth up to mixture $x_K = s_{2K-1} + s_{2K}$.

In our alternate batch-construction strategy, when constructing a batch of $K$ mixtures, rather than taking $2K$ speech signals and pairing them up individually (i.e. $x_k = s_{2k-1}+s_{2k}$), we use only $K$ speech signals and ensure every sample is paired twice for mixing:
\begin{align}
x_k = s_k + s_{k+1} \text{,}\label{eq:three_way}
\end{align}
with wraparound, i.e. $s_{K+1} \coloneq s_1$ and $s_0 \coloneq s_K$.
We call this ring mixing, illustrated in Figure~\ref{fig:ring_diagram}.
Note that each source $s_k$ is used in two different mixtures ($x_k$ and $x_{k-1}$).

These mixtures are passed through the network as normal, producing estimates $\hat{s}_{k;x_j}$, i.e. the estimate of $s_k$ produced from the mixture $x_j$.
Then, for each source, we use an additional SNR-like loss function called Signal-to-Consistency-Error Ratio~(SCER) loss to encourage the estimates of the same source from different mixtures to be the same:
\begin{align}
\ell_\text{SCER}(\hat{s}_{k;x_{k-1}},\hat{s}_{k;x_k};s_k) = - 10\log\frac{||s_k||^2}{||\hat{s}_{k;x_{k-1}} - \hat{s}_{k;x_k}||^2} \label{eq:scer_loss}\text{.}
\end{align}
Here again we note that mixtures $x_k$ and $x_{k-1}$ both were constructed using source $s_k$, and that we are essentially computing the log mean squared error between the estimates.
Although $s_k$ appears in the numerator, this is included only for interpretability as an SNR.
It has no impact on gradient computation, and so this loss is thus unsupervised.

We next return to the specific usage of data with noisy targets $s^\text{noisy}_k$ and evaluate the impact on this loss function.
Considering again estimates $\tilde{s}_k$ from the set $\mathcal{S}$, the effect on $\ell_\text{SCER}$ is:
\begin{align}
\lambda^\star & = \arg\min_\lambda \sum_k \ell_\text{SCER}(\tilde{s}_{k;x_{k-1}},\tilde{s}_{k;x_k};s^\text{noisy}_k) \\
& = \arg\min_\lambda \sum_k \log || \tilde{s}_{k;x_{k-1}} - \tilde{s}_{k;x_k} ||^2 \\
\begin{split}
& = \arg\min_\lambda \sum_k \log || \left[ s_k + \lambda(n_{k-1}+n_k)\right] \\[-7pt]
& \hspace{8.3em} - \left[s_k + \lambda(n_k+n_{k+1})\right] ||^2
\end{split} \\[2pt]
& = \arg\min_\lambda \sum_k \log || \lambda n_{k-1} - \lambda n_{k+1}||^2 \\
& = \arg\min_\lambda \sum_k \log \lambda^2 ||n_{k-1}-n_{k+1}||^2 \\
& = 0 \text{,}
\end{align}
which we see has the desired minimum at $\lambda = 0$, corresponding to fully-denoised separation.

In some sense, this approach resembles a self-supervised version of Noise2Noise~\cite{noise2noise}, in which image denoising is performed without clean-image supervision by training a network to predict \textit{one} noisy observation of an image from a \textit{different} noisy observation of the same image.
The assumption is that it is not possible to predict the noise in the training target, but it is possible to remove the noise in the training input.
As such, the ``best possible'' solution that the network is capable of producing is the denoised image, lacking both noise signals.

This parallels our analysis with the ``best possible'' set $\mathcal{S}$.
With ring mixing and SCER, the mechanism is that estimates of the same source from different mixtures will have different noise mixtures.
Each estimate's noise mixture cannot be predicted from the other estimate's input mixture, so the way SCER is maximized is to output only what can be estimated in common (the underlying clean speech).

Unfortunately, the sources are \textit{not} constrained to $\mathcal{S}$, and SCER is essentially a comparison between two arbitrary network outputs, which is very unstable and can lead to degenerate solutions.
As a result, we still use the conventional SI-SDR loss prior to computing the consistency loss, specifically using the \textit{estimate scaling} version of SI-SDR, retaining the rescaled estimates for use in SCER.
SI-SDR is also used for the necessary PIT permutation solving of sources.

Conceptually, the full procedure is simple: Use ring mixing as illustrated in Figure~\ref{sfig:ring_mixing}; compute SI-SDR loss, keeping the rescaled estimates; add the consistency loss SCER~\eqref{eq:scer_loss}, computed between the rescaled estimates of the same source from different mixtures.
Formally, the loss for a full batch is:
\begin{gather}
\mathcal{L} = \frac{1}{K}\sum_{k=1}^K\ell(\hat{s}_{k;x_{k-1}},\hat{s}_{k;x_k};s^\text{noisy}_k) \text{, where}\notag\\
\begin{split}
\ell(\hat{s}_{k;x_{k-1}},\hat{s}_{k;x_k};s^\text{noisy}_k) = \frac{1}{2}\sum_{j\in\{k-1, k\}}\ell_\text{SDR}(\beta_{k;x_j}\hat{s}_{k;x_j}; s^\text{noisy}_k) \\
+ \alpha\ell_\text{SCER}(\beta_{k;x_{k-1}}\hat{s}_{k;x_{k-1}},\beta_{k;x_k}\hat{s}_{k;x_k};s^\text{noisy}_k) \text{,}
\end{split} \notag \\
\text{for }\beta_{k;x_j}\text{ such that }s^\text{noisy}_k \perp s^\text{noisy}_k - \beta_{k;x_j}\hat{s}_{k;x_j} \text{.}\label{eq:loss}
\end{gather}
Note that a hyperparameter $\alpha$ has additionally been added to weight the SCER loss's contribution to the overall loss, which is typically set equal to $1$.

\section{Experimental Setup}
\label{sec:experimental_setup}
\subsection{Datasets}
\label{ssec:datasets}

The primary dataset we used for our experiments (which we refer to as WHAM!+) is the dataset described by Maciejewski et~al.~\cite{noisy_oracle}. It consists of the WHAM!~\cite{wham} dataset (wsj0-2mix~\cite{wsj2mix} with added noise), where each mixture is assigned an \textit{additional} WHAM! noise recording, such that each speech source in the mixture gets its own noise source.
The two speech recordings and two noise recording are summed to create a mixture $x = s_1+n_1+s_2+n_2$, at a configurable noisy-speech SNR.
The ground truth can be configured between the clean speech sources or the noisy speech sources (i.e. $s^\text{clean}_k = s_k$ and $s^\text{noisy}_k = s_k+n_k$), to study the effect of training with the different types of supervision while using identical mixtures.
Since this is an existing 2-speaker separation dataset, while generating a batch of size $K$, we sample only $K/2$ mixtures, but use the $K$ underlying sources to generate $K$ new mixtures following the ring-mixing procedure.

We also use the original wsj0-2mix test set, to study the speech separation and modeling capabilities of systems without any effects of noise.

Additionally, to evaluate our method on non-simulated noisy speech, we used the VoxCeleb~\cite{voxceleb1,voxceleb2} corpora, serving as a large source of in-the-wild naturally-noisy speech recordings.
As this is not a speech separation corpus, we use on-the-fly data generation, directly sampling $K$ single-speaker segments per batch to generate $K$ mixtures, again using ring mixing.
\vspace{-0.15cm}
\subsection{Model and Training Configuration}
\label{ssec:model}

We used a 4-block TF-GridNet~\cite{tfgridnet1,tfgridnet2} model, following the hyperparameters and training setup for noisy/reverberant separation and \SI{16}{kHz} audio described by Wang et~al.~\cite{tfgridnet2}.
The hyperparameter $\alpha$ was not tuned and was set to $1$, except in experiments probing its effect.

\subsection{Evaluation Metrics}
\label{ssec:evaluation_metrics}

Our primary metric is SI-SDR~\cite{sisdr} (and improvement SI-SDRi), used to evaluate system output against the noiseless ground truth.
This jointly measures separation and denoising, as is the goal of our method.

We also try to measure how much of the interfering sources remain in the estimate (e.g. $s_2$, $n_2$, and $n_1$ in $\hat{s}_1$, generalized across $k$'s to $s_\text{other}$, $n_\text{other}$, and $n_\text{self}$).
For this, we rescale the estimates to the clean speech and measure an ``occupancy'' metric, e.g.:
\begin{align}
occ._{n_1}(\hat{s}_1) \coloneq \frac{\langle\beta\hat{s}_1, n_1\rangle}{||n_1||^2},\text{for }\beta\text{ s.t. }s^\text{clean}_1\perp s^\text{clean}_1-\beta\hat{s}_1 \text{,}\hspace{-0.15em}\label{eq:occ}
\end{align}
where we attempt to measure how much of $n_1$ is contained in the estimate $\hat{s}_1$.
To aid in intuition: this is again designed to exploit the orthogonality of audio signals.
For example, if $\hat{s}_1 = s_1 + \lambda n_1$, $\langle\beta \hat{s}_1, n_1\rangle = \langle\lambda n_1, n_1\rangle = \lambda ||n_1||^2$, and so $occ._{n_1}(\hat{s}_1) = \lambda$.
In general, this can be roughly interpreted as a score from $0$ to $1$ reflecting the fraction of the interfering source remaining in the estimate, but the reader should beware that this metric can be greater than $1$ and less than $0$.

\section{Results and Discussion}
\label{sec:results_and_discussion}
\begin{table}[t]
\caption{Performance comparison of models trained using WHAM!+ at different levels of noise. The ``noisy'' and ``clean'' rows serve as performance floor/ceiling baselines, with the latter being trained with the ideal, clean-speech supervision. SI-SDR improvement on the noiseless wsj0-2mix dataset is included to measure the pure speech-modeling capabilities.}
\centering
\sisetup{
    reset-text-series = false,
    text-series-to-math = true,
    mode=text,
    round-mode=places,
    round-precision=2,
    table-number-alignment=center}
\begin{tabular}{ccS[table-format=2.1,round-precision=1]SSSS[table-format=2.1,round-precision=1]}
\toprule
 & & \multicolumn{4}{c}{WHAM!+} & {wsj0-2mix} \\ \cmidrule(lr){3-6} \cmidrule(lr){7-7}
 & & {\multirowcell{2.9}{SI-SDRi\\ {[dB] $\uparrow$}}} & \multicolumn{3}{c}{Occupancy \eqref{eq:occ} $\downarrow$} & {\multirowcell{2.9}{SI-SDRi\\ {[dB] $\uparrow$}}} \\ \cmidrule(lr){4-6}
{SNR} & {System} &  & {$s_\text{other}$} & {$n_\text{other}$} & \multicolumn{1}{c}{$n_\text{self}$} &  \\ \midrule
\multirow{3.2}{*}{\rotatebox[origin=c]{90}{\SI[round-precision=0]{20}{dB}}} & {noisy} & 11.372417115689565 & 0.033669969380091064 & 0.507639868626992 & 0.515665577719609 & 17.958570790213958 \\
 & {+SCER} & 13.017032492410392 & 0.026787022975409247 & 0.3547945469021797 & 0.35926104825735095 & 17.82433085533119 \\ \cmidrule{2-7}
 & {clean} & 14.592000816696276 & 0.023461214813646318 & 0.13894642659276724 & 0.13798334216326474 & 17.956240072287105 \\ \midrule
\multirow{3.2}{*}{\rotatebox[origin=c]{90}{\SI[round-precision=0]{10}{dB}}} & {noisy} & 11.388065564696628 & 0.03935491480053558 & 0.5090364592174689 & 0.525570049415032 & 16.270974507721093 \\
 & {+SCER} & 13.305148180424808 & 0.030171230987199427 & 0.22839681059867142 & 0.23392459633449714 & 14.779582107343508 \\ \cmidrule{2-7}
 & {clean} & 15.075720807437737 & 0.02201361977920654 & 0.09256335403583944 & 0.09102476127600918 & 16.737218777700743 \\ \midrule
\multirow{3.2}{*}{\rotatebox[origin=c]{90}{\SI[round-precision=0]{00}{dB}}} & {noisy} & 9.34379058658968 & 0.1050722776701053 & 0.5602147264877955 & 0.5725107676486174 & 10.394668693614996 \\
 & {+SCER} & 10.462091376074978 & 0.09795933484658599 & 0.41292648560305434 & 0.42233548501133916 & 10.424952989461017 \\ \cmidrule{2-7}
 & {clean} & 13.015283510850074 & 0.05554035065963399 & 0.1051797635583207 & 0.10430348678616186 & 12.465151632709667 \\
\bottomrule
\end{tabular}
\vspace{-7pt}
\label{tab:full_results}
\end{table}

The results of our initial experiments are in Table~\ref{tab:full_results}, in which we trained and evaluated systems on three separate noise levels of WHAM!+: \SI{20}{dB} for low, \SI{0}{dB} for high, and \SI{10}{dB} for a roughly ``typical'' amount of noise.
The ``noisy'' row represents the SI-SDR baseline trained with the $s^\text{noisy}_k$ targets, while the ``clean'' row is the SI-SDR baseline trained with the $s^\text{clean}_k$ targets, serving as the performance ceiling for the mixtures.
As no prior single-channel separation method denoises without clean-speech supervision, these systems are our only points of comparison.
For fairest comparison, both baselines were trained with ring mixing as well (e.g. ``noisy'' corresponds to $\alpha=0$) to demonstrate the effects of SCER~\eqref{eq:scer_loss}, noting that we have observed no meaningful performance differences between baseline models trained with and without ring mixing.

In all cases, including SCER while training on noisy-target data increases SI-SDR improvement (SI-SDRi) on WHAM!+, by \SI{1.2}-\SI{1.9}{dB}, closing about half the gap to the ideal, clean-speech supervision.

The results become more interesting when investigating the presence of the underlying signals, measured by our occupancy metric \eqref{eq:occ}.
The presence of the non-target, interfering speech $s_\text{other}$ is low in all models, showing little impact from the supervision or loss function (though SCER seems more likely to help than hurt), suggesting that the issues discussed in this paper are largely issues of noise, not separation of speech.

In terms of the presence of noise in the estimates ($occ._{n_\text{other}}$ and $occ._{n_\text{self}}$), SCER appears to have a strong impact, reducing the amplitude of both noises by roughly half.
The occupancy metrics tell an even richer story, though:
The fact that $occ._{n_\text{other}}$ and $occ._{n_\text{self}}$ are roughly equal in all cases (despite the loss functions being minimized at $occ._{n_\text{other}}=0$ and $occ._{n_\text{self}}=1$) is evidence that the network struggles to separate noise, and our set $\mathcal{S}$~\eqref{eq:set} is a reasonable assumption.
And, the noisy-speech supervised baseline systems have $occ._{n_*}$ values close to $0.5$, which matches the undesirable $\lambda^*=0.5$ minimum derived in Section~\ref{ssec:issue_with_conventional_supervision} that results from $\mathcal{S}$ with SI-SDR loss.

One downside we observed is that when evaluating on the noiseless wsj0-2mix condition, SCER systems show some degradation, suggesting decreased fidelity in modeling speech.

\begin{figure}[t]
\centering
\subfigure[Occupancy of Interfering Sources in \SI{10}{dB} WHAM!+]{
\input{figures/occupancy_plot.pgf}
\label{fig:occ_plot}
}
\vspace{1pt}

\subfigure[SI-SDR in Noiseless and Noisy Conditions]{
\input{figures/sisdr_plot.pgf}
\label{fig:sisdr_plot}
}

\caption{Validation metrics over the first \SI{100}{k} training steps on \SI{10}{dB} WHAM!+, comparing SCER systems at various $\alpha$ values trained with $s^\text{noisy}$ supervision to the baseline systems trained with both $s^\text{noisy}$ and $s^\text{clean}$ supervision.}
\vspace{-7pt}
\label{fig:alpha_plot}
\end{figure}

To demonstrate the effect of the mixing weight $\alpha$ of SCER, in Figure~\ref{fig:alpha_plot} we show plots of various validation-set metrics of systems over the course of training the \SI{10}{dB} WHAM!+ systems.
First, looking at the occupancy metrics in Figure~\ref{fig:occ_plot}, we see that at all points in training, $occ._{s_\text{other}}$ is nearly identical for all systems, again confirming that neither the supervision nor SCER loss seem to have any significant effect (good or bad) on the suppression of interfering speech.

In the occupancy metrics for noise, we again see that in all cases the curves for $occ._{n_\text{other}}$ and $occ._{n_\text{self}}$ are nearly identical (more evidence of noise inseparability).
And we see the convergence of noisy-target SI-SDR system to the $\lambda^*=0.5$ minimum.
In terms of the effect of $\alpha$: as the SCER loss is included and its weight is increased, we see improved noise suppression, with the $\alpha=2.0$ system matching the clean-speech supervised baseline in noise removal.

\begin{figure}[t]
\centering
\import{figures}{specs.pgf}
\caption{Example spectrograms from the \SI{10}{dB} validation set, showing an input mixture as well as outputs from the baseline with $s^\text{clean}$ supervision and the $s^\text{noisy}$-supervised SCER system with $\alpha=2.0$.}
\vspace{-12pt}
\label{fig:specs}
\end{figure}

Although the occupancy metrics suggest SCER is purely beneficial and a large $\alpha$ is best, the tradeoffs are demonstrated in the overall SI-SDR performance in Figure~\ref{fig:sisdr_plot}.
Again, there is degradation in the clean wsj0-2mix condition, with the models appearing to fall into local minima, terminating earlier as $\alpha$ increases, leading to reduced output speech fidelity.

Qualitatively, we observe that the errors are typical of weaker systems rather than showing any pathological effect, as demonstrated via spectrograms in Figure~\ref{fig:specs}, where we show an example mixture as well as the output of the clean-speech target SI-SDR baseline and the SCER $\alpha=2.0$ system, which suffered this effect the worst.
First, focusing on the black box regions, we can see that the interfering speech signal (visible in the mixture) does not appear to be present in either system output, demonstrating that the SCER system does not leak interfering speech (as reported by $occ._{s_\text{other}}$).
The red box regions highlight some types of degradations that \textit{are} seen in the SCER system: a loss of detail, particularly in high-frequency and low-amplitude areas.

But, returning to SI-SDR on WHAM!+, in Figure~\ref{fig:sisdr_plot}: though this early termination effect persists, the denoising improvements outweigh it for many systems, resulting in overall SI-SDR improvements over the baseline.

A key highlight from Figure~\ref{fig:alpha_plot} is that SCER seems to be a very stable auxiliary loss.
The weight $\alpha$ does \textit{not} need to be tuned for convergence, with a wide variety of values still outperforming the baseline in pure signal reconstruction.
Rather, it serves for design decisions as to whether speech fidelity or denoising are more important, essentially adjusting the tradeoff in performance between the two. 
This finding motivated our decision to not tune $\alpha$ for our other experiments.

\begin{table}[t]
\caption{Denoising separation SI-SDRi~[dB] comparison on WHAM!+ dataset, demonstrating generalization capabilities. Table \ref{tab:full_results} results are repeated for ease of comparison, where each model was trained on the evaluation-set SNR. The mixed-SNR model was trained on the combination of these datasets. Evaluation on WHAM! is also included to aid in contextualizing results to existing literature.}
\centering
\sisetup{
    reset-text-series = false,
    text-series-to-math = true,
    mode=text,
    round-mode=places,
    round-precision=1,
    table-format=2.1,
    table-number-alignment=center}
\begin{tabular}{ccSSSS}
\toprule
{\multirowcell{2.95}{Training\\ Dataset}} & & \multicolumn{3}{c}{WHAM!+} & \\ \cmidrule(lr){3-5}
 & System & {\SI[round-precision=0]{20}{dB}} & {\SI[round-precision=0]{10}{dB}} & {\SI[round-precision=0]{0}{dB}} & {WHAM!} \\ \midrule
\multirow{1.5}{*}{\multirowcell{1.0}{WHAM!+ \\ matched SNR \\ (Table \ref{tab:full_results})}} & noisy & 11.372417115689565 & 11.388065564696628 & 9.34379058658968 &   \\
 & +SCER & 13.017032492410392 & 13.305148180424808 & 10.462091376074978 &   \\ \cmidrule{2-5}
 & clean & 14.592000816696276 & 15.075720807437737 & 13.015283510850074 &   \\ \midrule
\multirow{2.1}{*}{VoxCeleb} & noisy & 13.614358436197216 & 9.610544893923322 & 4.952229144483883 & 5.6177794205664044 \\
 & +SCER & 14.562049457679056 & 11.335200187223915 & 6.357188863876426 & 7.227793507606955 \\ \midrule
\multirow{2.5}{*}{\multirowcell{1.0}{WHAM!+ \\ mixed SNR}} & noisy & 14.91263288585766 & 10.942770552888154 & 6.477337684050114 & 7.136366794516827 \\
 & +SCER & 15.398075823086819 & 13.536303522034238 & 10.284570850951287 & 10.872234510912909 \\ \cmidrule{2-6}
 & clean & 16.08749541371799 & 14.736410978391408 & 14.254709456339167 & 14.594242590548781 \\
\bottomrule
\end{tabular}
\vspace{-7pt}
\label{tab:voxceleb_results}
\end{table}

Our final experiments investigate the generalizability of our approach, the results of which are in Table~\ref{tab:voxceleb_results}.
We first call attention to evaluating the proposed method while training on VoxCeleb, where the supervision comes from real noisy recordings rather than simulated noisy recordings (though still using WHAM! test sets to enable measurement of denoising).
In all conditions, the VoxCeleb model trained with SCER shows small but consistent improvements in denoising speech separation, indicating the approach is effective while using real data as well.
The small gains and stronger performance on the low-noise \SI{20}{dB} condition compared to WHAM!+ models is consistent with the fact that VoxCeleb's celebrity interviews are not exceptionally noisy.

Particularly interesting results emerged from training on the combination of the three chosen WHAM!+ conditions, mimicking more-realistic variable-SNR data.
Focusing on the baselines, the regular noisy-target SI-SDR system outperforms the matched-condition baseline system in \SI{20}{dB}, indicating that adding additional lower-SNR training data helps performance in high-SNR conditions, but performance actually \textit{significantly degrades} in the \SI{0}{dB} condition.
This suggests that well-trained separation systems cannot be adapted to noisy conditions through merely adding mixtures of in-domain noisy data.
Fortunately, adding in SCER loss is still universally effective, and mostly undoes the aforementioned damage of mixed-condition training in the high-noise scenario.

Further investigation of this is left to future work, but we theorize that this is a consequence of SI-SDR actively incentivizing including noise in the estimate.
Since the optimum is a specific fraction of the total noise, and the amount of noise varies between samples, the optimization target is accordingly inconsistent across the conditions present in the training data.
The denoised optimum of SCER helps resolve this target inconsistency, aiding generalization, and potentially is a key to extending synthetic-trained systems to real-world conditions.

\section{Conclusion}
\label{sec:conclusion}
In this work, we have demonstrated that using SI-SDR loss while training speech separation systems using mixtures of naturally-noisy speech results in an undesirable optimum, a potential contributor to the limited successes of separation in practical environments, where in-domain speech training data often includes noise.
To address this, we have introduced the ring mixing batch construction strategy and SCER auxiliary loss, resulting in significant improvements in denoising (despite using no paired noisy- and clean-speech recordings) and better generalization of systems.
We are optimistic that this approach serves as a promising first step towards extending the capabilities of denoising separation systems trained on fully-synthetic mixtures to real-world recordings featuring overlapping speech.

\bibliographystyle{IEEEtran}
\bibliography{mybib}

\end{document}